\pdfoutput=1

\documentclass[prl,superscriptaddress,amsmath,amssymb,twocolumn]{revtex4-1}

\usepackage{graphicx,bm,amsmath}
\usepackage[usenames,dvipsnames]{xcolor}
\usepackage[colorlinks,bookmarks=false,citecolor=NavyBlue,linkcolor=Red,urlcolor=blue]{hyperref}

\usepackage[bbgreekl]{mathbbol}
\usepackage{xcolor}

\def\doi{http://dx.doi.org/}

\newcommand{\be}{\begin{equation}}
\newcommand{\ee}{\end{equation}}
\newcommand{\bea}{\begin{eqnarray}}
\newcommand{\eea}{\end{eqnarray}}

\newcommand{\up}{\uparrow}
\newcommand{\down}{\downarrow}

\def\nn{\nonumber\\}

\def\pp{\upsilon}
\def\nn{\mathfrak{n}}
\def\sgn{\operatorname{sgn}}

\newcommand{\titleinfo}{Domain-wall melting in the spin-$1/2$ XXZ spin chain:\\
emergent Luttinger liquid with fractal quasi-particle charge}

\begin{document}

\title{\titleinfo}
\author{Mario Collura}
\affiliation{SISSA, Via Bonomea 265, 34136 Trieste, Italy}
\author{Andrea De Luca}
\affiliation{Laboratoire de Physique Th\'eorique et Mod\'elisation (CNRS UMR 8089),
Universit\'e de Cergy-Pontoise, F-95302 Cergy-Pontoise, France}
\author{Pasquale Calabrese}
\affiliation{SISSA, Via Bonomea 265, 34136 Trieste, Italy}
\affiliation{INFN, Via Bonomea 265, 34136 Trieste, Italy}
\affiliation{International Centre for Theoretical Physics (ICTP), I-34151, Trieste, Italy }
\author{J\'er\^ome Dubail}
\affiliation{Laboratoire de Physique et Chimie Th\'eoriques, CNRS, UMR 7019,
Universit\`e de Lorraine, 54506 Vandoeuvre-les-Nancy, France}
    
\begin{abstract}
In spin chains with local unitary evolution preserving the magnetization $S^{\rm z}$, the domain-wall state $\left| \dots \uparrow \uparrow \uparrow \uparrow \uparrow \downarrow \downarrow \downarrow \downarrow \downarrow  \dots \right>$ typically ``melts''. At large times, a non-trivial magnetization profile develops in an expanding region around the initial position of the domain-wall. For non-integrable dynamics the melting is diffusive, with entropy production within a melted region of size $\sqrt{t}$. In contrast, when the evolution is integrable, ballistic transport dominates and results in a melted region growing linearly in time, with no extensive entropy production: the spin chain remains locally in states of zero entropy at any time. Here we show that, for the integrable spin-$1/2$ XXZ chain, low-energy quantum fluctuations in the melted region give rise to an emergent Luttinger liquid which, remarkably, differs from the equilibrium one. 
The striking feature of this emergent Luttinger liquid is its quasi-particle charge (or Luttinger parameter $K$) which acquires a fractal dependence on the XXZ chain anisotropy parameter $\Delta$. 
\end{abstract}

\maketitle

\paragraph{\it Introduction. ---}\label{sec:intro} The phenomenon of domain-wall (DW) melting in quantum magnetism is a simple example of quantum many-body dynamics. 
It has a long history in the context of quantum spin chains, dating back to early experimental work on ${\rm Co Cl}_2 \cdot 2 {\rm H}_2 {\rm O}$ chains \cite{torrance1969excitation} which provided the initial motivation for many subsequent theoretical developments. Those include studies of the dynamical stability of domain walls \cite{gochev1977spin,gochev1983contribution,yuan2007domain}, exact calculations of magnetization profiles in free fermion chains \cite{antal1999transport,hunyadi2004dynamic,platini2005scaling,platini2007relaxation,  NESSf5, NESSf6, viti2016inhomogeneous}, approximate and numerical analysis 
both in integrable and non-integrable spin chains~\cite{gobert2005real,chl-08,jesenko2011finite,zauner2012time,halimeh2014domain,alba2014entanglement,hauschild2016domain, NESSnum1,NESSnum2,NESSnum3, DVMR14,bernarddoyon2016, NESSf7, NESSf11,rakovszky2019entanglement,bulchandani2019subdiffusive}.

On the analytical side, the 2016 discovery of a hydrodynamic approach to quantum integrable systems~\cite{bertini2016transport,castro2016emergent}, now dubbed  \textit{Generalized Hydrodynamics} (GHD), has provided the ultimate analytical tool to analyze inhomogeneous dynamics of integrable systems~\cite{doytaka2017,DDKY17,bulchandani2018bethe,bastianelloalba2019,ms-20}, even in the classical context~\cite{BDWY17,ds-17,dsy-17}. Its application to the domain-wall melting in integrable spin chains has been particularly effective, providing the exact magnetization profile at large time for the XXZ chain~\cite{bib:cdv18} (see below).

Much effort has been spent to extend such a powerful method and include diffusive or superdiffusive
effects~\cite{ghddiff, vasseurdiff,milosz2019,superdiffusive2018,doyonsuper}, 
non-ballistic phenomena~\cite{nonball2017} and integrability breaking~\cite{kapitza2019, friedman2019, flux2019}  
so as to codify correlations~\cite{quasilongrange, fagotti2017, fagotti2019, brun2018, cftbreathing2019} 
and entanglement~\cite{BFPC18, alba2019, BAF19, bastiadubail}.
Indeed, due to its own coarse-grained nature, GHD in its original form 
cannot account for entanglement generation and quantum correlation spreading following the quantum unitary evolution. 
As a matter of fact, an uncorrelated initial state -- e.g. a product state with zero entanglement -- 
does develop entanglement when it evolves under a nontrivial unitary evolution. 
Recently, a low-energy description in terms of multi-component Luttinger liquids (LL)~\cite{fokkema2014split,eliensjs,correlations2016vlijm} has been put forward~\cite{ruggiero2019}; such refined ``quantum" adaptation of the GHD has been tested for integrable quantum gases~\cite{ruggiero2019}.
Here we further develop this intuition and explore the non-equilibrium dynamics from a domain-wall (DW) state
in the XXZ spin-$1/2$ chain, a genuinely interacting integrable model.
Despite the simple structure of the initial state, the 
dynamics is highly nontrivial \cite{bib:cdv18}; 
interestingly, the emerging local quasi-stationary state (LQSS)~\cite{bertinifagotti2016, bertini2016transport, Bas_Deluca_defhop, Bas_Deluca_defising} admits a description in terms of two species of particles, each supporting a single Fermi point. In the spirit of the quantum GHD picture~\cite{ruggiero2019}, 
we show that the quantum fluctuations in the LQSS can be exactly encoded in a 
LL, whose Luttinger parameter is nontrivial and differs from the standard low-energy equilibrium one~\cite{giamarchi2004one,sirker2005} 
governing the transport at low temperature \cite{bpc-18,bp-18}.

\paragraph{Model and GHD solution of domain-wall. ---}\label{sec:model_setup}
We consider the unitary dynamics generated by the
one-dimensional spin-$1/2$ XXZ Hamiltonian
\be\label{eq:H}
H =  \sum_{x = -\infty}^{\infty} 
S^{{\rm x}}_{x}S^{{\rm x}}_{x+1}
+ S^{{\rm y}}_{x}S^{{\rm y}}_{x+1}
+ \Delta S^{{\rm z}}_{x}S^{{\rm z}}_{x+1} ,
\ee
where $S^{\alpha}_{x}$ are spin-$1/2$ operators acting on the site $x$.
We focus on the regime $-1< \Delta < 1$ which exhibits ballistic transport~\cite{PrIl13, ID117, bib:cdv18}. 
[It is known that for $|\Delta| > 1$ the domain wall does not melt, see e.g. the energetic argument given in Ref.~\cite{bib:mmk17,mpp-19}, and $\Delta = 1$  is pathological~\cite{LZP17, superdiffusive2018}.] Moreover we focus on the `rational case' where the anisotropy $\Delta$ is parameterized as 
\begin{equation}
	\label{eq:Delta}
\Delta = \cos(\gamma), \quad \gamma = \pi  Q/P,
\end{equation}
where $Q$ and $P$ are two co-prime integers with $1\leq Q < P$.
The initial state is the classical DW state in the $\rm z$ direction, 
$| {\rm DW} \rangle = |\cdots \up\up\up\down\down\down\cdots\rangle$ and
it undergoes unitary evolution generated by the Hamiltonian (\ref{eq:H}), i.e.
$|\Psi (t) \rangle = e^{-i t H} | {\rm DW} \rangle$.
In Ref.~\cite{bib:cdv18}, the exact large-time magnetization profile was calculated using GHD, as we now briefly recall. 
GHD is a hydrodynamic approach valid at large distances and on long time scales, where the local state of the system in a space-time cell $[x,x+dx] \times [t, t+dt]$ is represented by a Fermi filling factor $\vartheta_{j} (x,t,\lambda) \in [0,1]$ for each species $j$ of quasi-particles, or string, with rapidity $\lambda$ \cite{bertini2016transport}. For the XXZ chain in the rational case, the index $j$ is an integer ranging from $1$ to $\ell = \sum_{k=1}^\delta \nu_k$, where the ratio $Q/P$ has been represented as a finite continued fraction $Q/P=\frac{1}{\nu_1 + \frac{1}{\nu_2 + \ldots}}$
of length $\delta$~\cite{takahashi}. The GHD equations then read~\cite{bertini2016transport}
\begin{subequations}
	\label{eq:GHD}
\begin{eqnarray}
	&& \partial_t \vartheta_{j} (\lambda; x, t) + v^{\rm eff}_j (\lambda) \partial_x \vartheta_{j} (\lambda; x,t) = 0 ,\\
	&& v^{\rm eff}_j (\lambda) \, = \, \frac{\partial_\lambda \varepsilon (\lambda)}{\partial_\lambda p (\lambda)} ,
\end{eqnarray}
\end{subequations}
where $\varepsilon_j(\lambda)$ and $p_j(\lambda)$ are the energy and momentum of a quasi-particle of species $j$ with rapidity $\lambda$. Their explicit expressions is not essential for us and can be found in \cite{Note1}.
For general initial states, the GHD equations have to be solved numerically~\cite{bertini2016transport,castro2016emergent}, but for the special case of the DW initial state 
they admit an analytical solution~\cite{bib:cdv18}. 
This stems from two remarkable observations: 
(i) in the initial state, all filling factors are identically zero or one, i.e. $\vartheta_j(\lambda; x, t=0) = 1$ for $j \in \{\ell-1, \ell\}$ and $\lambda<0$ and vanishes otherwise;
(ii) in those local macrostates the effective velocity takes a very simple form independent of space and time
\be
v^{\rm eff}_{j}(\lambda) = \frac{\sin(\pi Q/P)}{\sin(\pi/P)} \sin(\sigma_{j} p_{j}(\lambda)), \quad j \in\{ \ell, \ell-1\},
\ee
where $\sigma_{j}  = \sgn(p_j'(0))$ is the ``sign'' of the string, defined so that
$\sigma_{j} p_{j} (\lambda)$ is a strictly increasing function of $\lambda \in [-\pi/P,\pi/P]$.
Then the equation (\ref{eq:GHD}a) is easily solved~\cite{bib:cdv18},
\begin{equation}
\label{LQSSDW}
	\vartheta_{j} (\lambda; x, t) = \left\{ \begin{array}{ccc}
		1 &{\rm if}& x/t > v^{\rm eff}_j(\lambda) \quad {\rm and} \quad j \in \{ \ell-1,\ell \} \\
		0 && {\rm otherwise} .
	\end{array} \right.
\end{equation}
The local macrostate parameterized by the filling factor $\vartheta_j(x,t,\lambda)$ thus depends only on the ratio $\zeta=x/t$; this is of course expected since the problem of domain-wall melting is a particular case of the more general Riemann problem in hydrodynamics~\cite{riemann1860}. The velocity does not depend explicitly on $x/t$, but the effect of the interactions is such that the light-cone is shrunk as $x/t \in [-\sin(\pi P/Q),\sin(\pi P/Q)]$. This led to analytic formulae for the profiles of the stationary magnetization and spin current~\cite{bib:cdv18} (see also \cite{Note1} for a short summary).

\paragraph{Effective LL for quantum fluctuations in the melted region --- } The goal of this paper is to investigate quantities that go beyond the classical Euler-scale GHD equations (\ref{eq:GHD}a-b), such as the bipartite entanglement entropy or the quantum fluctuations of the magnetization. This requires to describe quantum fluctuations around the GHD solution.
The dynamics from the DW state is fully characterized 
by the last two strings. In particular, for any ray $\zeta = x/t$, each of them has one single Fermi point $\lambda^\ast$ where the filling factor $\vartheta(\lambda^\ast; x,t)$ jumps from $0$ to $1$. Following the logic of Ref.~\cite{ruggiero2019}, this leads to an effective inhomogeneous LL with action
\begin{equation}
	\mathcal{S} \, = \, \frac{1}{8 \pi} \int \frac{dx dt}{K} g^{ab} (\partial_a h) (\partial_b h),
\end{equation}
where $a,b = x,t$ and $h(x,t)$ is the height field related to the fluctuations of the local 
magnetization as $S^{\rm z}_x - \left< S^{\rm z}_x \right> = \frac{1}{2\pi} \partial_x h$, and the metric is
$
ds^2 = \left( \frac{x}{t} dt -dx \right)^2
$~\cite{dubail2017conformal}.
As in equilibrium, the Luttinger parameter is the square of the quasi-particle charge~\cite{le2008charge}, or dressed magnetization, evaluated on any of the two Fermi points~\cite{bastiadubail,
eliensjs}. Moreover, the requirement of a field theory without chiral anomaly implies that the two strings give the same Luttinger parameter. This is confirmed by the explicit calculation which leads to~\cite{Note1}
\begin{equation}
	K = [n(\lambda^\ast)]^2 = \frac{P^2}{4}.
	\label{KP}
\end{equation}
Remarkably, since $K$ depends only on the denominator of $\gamma/\pi$, 
it exhibits a fractal (i.e. nowhere continuous) dependence on the anisotropy parameter $\Delta = \cos (\pi P/Q)$.
Also, since the Luttinger parameter does not depend on $x$ and $t$, this is a particularly simple version of an inhomogeneous LL where conformal invariance is not broken~\cite{dubail2017emergence}. 
Consequently, the correlation functions of primary fields $\phi_1$, \dots, $\phi_n$ with scaling dimensions $\Delta_1$, \dots, $\Delta_n$ obey the scaling relation
\begin{eqnarray}
	\label{eq:scaling}
	&& \left< \phi_1(x_1,t_1) \dots \phi_n(x_n,t_n) \right>  \, = \\
\nonumber	&& \qquad \prod_{i=1}^n (\tau/t_i)^{\Delta_i} \times \left< \phi_1( \tau x_1/t_1 ,\tau ) \dots \phi_n(\tau x_n/t_n,\tau) \right>,
\end{eqnarray}
for any fixed $\tau$. In other words, all correlation functions can be expressed in terms of equal-time correlations at some fixed time $\tau$. However, we cannot yet fully determine the correlation functions at time $\tau$, as this would require an exact lattice calculation. In the free fermion case $\Delta = 0$, such a calculation is possible~\cite{dubail2017conformal} using a clever Euclidean-time regularization which connects to a two-dimensional inhomogeneous statistical problem~\cite{allegra2016inhomogeneous}. In principle, a similar regularization should also be possible for $\Delta \neq 0$~\cite{granet2019inhomogeneous}, but presently we do not know how to do this calculation. Nevertheless, the scaling relation (\ref{eq:scaling}) is sufficient to derive a number of non-trivial results about 
quantum correlations in the long-time behavior of the system which we  summarise in the following.

\begin{figure}[t!]
\begin{center}
\includegraphics[width=0.47\textwidth]{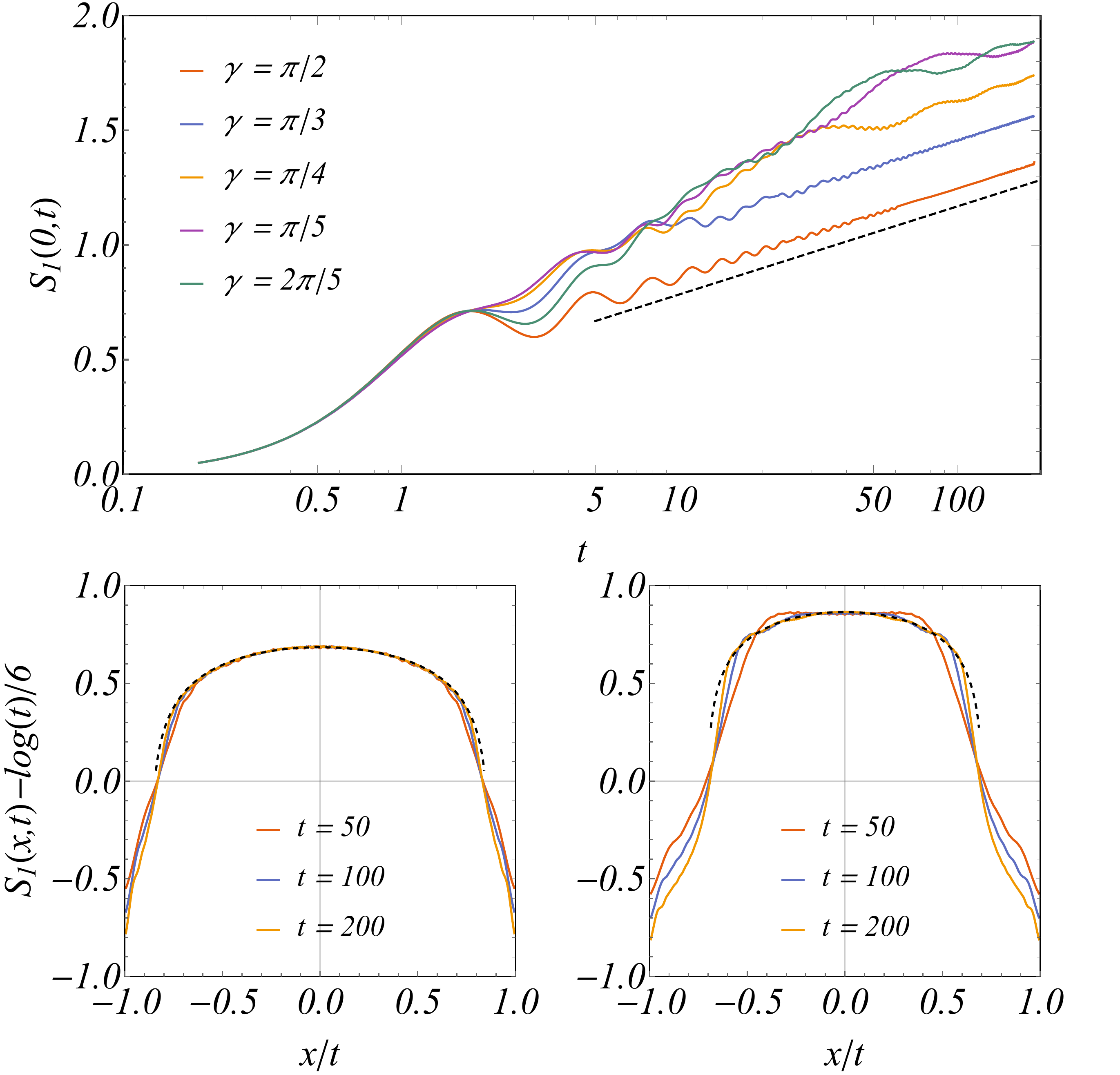}
\caption{\label{fig:St_NESS}\label{fig:Sz}
{\bf (top panel)} The time evolution of the entanglement entropy between the two halves of the system,
namely $[-L/2,-1]$ and $[0,L/2-1]$ is plotted in log-linear scale. 
Different colors represent the TEDB simulations for different values of the anisotropy $\Delta$. 
The black dashed line is a guide for the eyes, representing the asymptotic leading behavior $\sim \log(t)/6$.
{\bf (bottom panels)} 
The profiles of the entanglement entropy as a function of the ray $x/t$ for different times $t$,
and $\gamma = \pi/3$ (left), $\pi/4$ (right). 
The black dashed lines are the phenomenological approximation (\ref{eq:Sz}).
}
\end{center}
\end{figure}

\paragraph{Entanglement entropy. ---} 
We consider the  entanglement entropy for the bipartition $A = (-\infty, x-1]$ and $B= [x,\infty)$, i.e.
$
S_\alpha(x,t) = \frac{1}{1-\alpha} \log ( {\rm Tr} [\rho^\alpha_A(t)] ),
$
where 
$
\rho_A(t) = {\rm Tr}_{B} |\Psi(t)\rangle \langle\Psi(t)|,
$
is the reduced density matrix of subsystem $A$. In the effective Luttinger liquid description, the trace ${\rm Tr} [\rho^\alpha_A(t)]$ can be obtained as the expectation value of a twist field $\left< \phi (x,t) \right>$ in a theory with replicas~\cite{calacardy2009}. 
The twist field $\phi$ is a primary operator with scaling dimension $\Delta = \frac{1}{12} (\alpha- 1/\alpha)$; therefore, the scaling relation (\ref{eq:scaling}) leads to 
\begin{eqnarray}
	\label{eq:Salpha}
\nonumber	S_\alpha(x,t) &=& \frac{1}{1-\alpha} \log ( \left< \phi (x,t)\right>/\epsilon ) \\
\nonumber	 &=& \frac{1}{1-\alpha} \log \left( (\tau / t)^{\frac{1}{24} (\alpha- \frac{1}{\alpha})} \left<  \phi (\tau x/t, \tau) \right>/\epsilon \right) \\
	 &=& \frac{1}{12} (1+\frac{1}{\alpha}) \log (t/\tau) +  f_\alpha(x/t) ,
\end{eqnarray}
where $\epsilon$ is a UV length scale which appears when one takes the continuum limit of the lattice model. While for homogeneous systems $\epsilon$ is simply a constant, in inhomogeneous setups it depends on the LQSS; in particular, in our setup it can depend on the ratio $\zeta=x/t$.
We thus 
set $f_\alpha(\zeta) = \frac{1}{1-\alpha} \log \left[\left<  \phi (\tau \zeta, \tau) \right>/\epsilon(\zeta)\right]$, which is an unknown function of $\zeta$.

When evaluated at fixed $x$ and in the limit $t \rightarrow \infty$, the entanglement entropy, therefore, exhibits a leading logarithmic universal behavior. 
For the von Neumann entropy ($\alpha \rightarrow 1$) this gives
\be\label{eq:St_NESS}
S_{1}(x,t) \underset{\underset{x \; {\rm fixed}}{t \rightarrow \infty}}{\sim} \frac{1}{6} \log t  + c_0(\Delta)+o(1) ,
\ee
where the sub-leading term $c_0(\Delta)$ eventually depend on $\Delta$. 
In the top-left panel of Figure~\ref{fig:St_NESS} we show the entanglement entropy $S_{1}(x=0,t)$.  
The perfect logarithmic behavior, with a prefactor independent of the value of the 
anisotropy $\Delta$, and compatible with the predicted value $1/6$, nicely confirms the expectations from the LL description of the melted region.
The equilibration occurs much faster for smaller values of the denominator $P$;
for $Q/P = 1/2$ and $1/3$, the large-time stationary regime has been reached at accessible times and the entanglement entropy 
perfectly matches the logarithmic growth (\ref{eq:St_NESS}).
Also for $Q/P = 1/4$, the approach to the LL regime is  evident, despite the presence of slowly decaying oscillations. 
Remarkably, for $Q/P = 2/5$, although the value of the anisotropy is relatively small, $\Delta \simeq 0.309$,
the relaxation toward the asymptotic regime is very slow.
Notice finally that for $Q/P = 1/5$ and $2/5$, 
the entanglement entropy approaches the LL asymptotics oscillating around the same curve, implying that the non-universal additive constant $c_{0}(\Delta)$ 
is the same for the two cases. 
This observation suggests that $c_{0}(\Delta)$ may depend only on $P$ (i.e. $K$), although we do not have a  field-theory explanation supporting this.

Next, we study the entanglement entropy $S_1(x,t)$ for fixed $\zeta=x/t$ when $t\rightarrow \infty$. 
As explained above, the profile function $f_{1}(\zeta)$ is hard to compute because it gets contributions both from the field theory and from the lattice regularization. 
Nevertheless, we can calculate it numerically, as shown in Fig.~\ref{fig:St_NESS}. 
The numerical results are well-approximated by the phenomenological formula
\be\label{eq:Sz}
f_1(\zeta) \simeq 
\frac{1}{6}\left(1 + \frac{1}{P} \right)\log\left[  1 - \left(\frac{\zeta}{\sin(\pi P/Q)}\right)^{2} \right] ,
\ee
designed such that for $Q/P=1/2$ it reproduces the exact result  for $\Delta = 0$~\cite{dubail2017conformal}. 
We do not have a theoretical justification of Eq. \eqref{eq:Sz}, nonetheless it undeniably provides a rather good approximation.

In Fig.~\ref{fig:Sz} (left-bottom panel), we show the  profile of the entanglement entropy 
for $Q/P =1/3$ and different times (larger than $50$ for which system is in the LQSS from the measure of $S_1(0,t)$). The profile is well approximated by Eq. (\ref{eq:Sz}), excepts from tiny regions close to the light-cone $x/t \simeq \pm \sin(\gamma)$.
For $Q/P=1/4$ (right-bottom panel in Figure~\ref{fig:Sz}) the largest time accessible by
time-evolving block decimation (TEBD)~\cite{TEBD} simulations is not sufficient to observe a complete relaxation 
to the large-time stationary behavior. For this reason, oscillations on top of the asymptotic profile
are present, but the agreement with Eq. (\ref{eq:Sz}) is fairly good.

\paragraph{Full counting statistics. ---} 
We now turn to the fluctuations of the magnetization $M_{[x_1,x_2]} = \sum_{x=x_1}^{x_2} S^{{\rm z}}_{x} $ in an interval $[x_1,x_2]$ inside the melted region. 
The generating function of the cumulants is
\begin{multline}
\label{eq:fullcounting}
F_{[x_1,x_2]} ( \lambda, t)  = \langle \exp \left(-i \lambda M_{[x_1,x_2]} \right) \rangle_t =\\=
\langle e^{ i \frac{\lambda}{2\pi} h(x_1,t) } e^{ -i \frac{\lambda}{2\pi} h(x_2,t) }  \rangle
\, ,
\end{multline}
where in the second line we used that the local magnetization 
is related to the height field as $S^{\rm z}_x - \langle S^{\rm z}_x \rangle = \frac{1}{2\pi} \partial_x h$.
We are interested in the case of an interval $[x-l/2,x+l/2]$ with fixed length $l \gg 1$,
in limit of large time $t \gg l$, keeping $\zeta=x/t$ fixed. 
Since $e^{i \alpha h(x,t)}$ is a primary field with scaling dimension $\Delta = \alpha^2 K$, 
for large $l$ the generating function behaves as
\be\label{eq:fullcountingscaling}
F_{[x-l/2,x+l/2]} ( \lambda, t) \simeq \left(\frac{l}{\epsilon' (x/t)}\right)^{-\frac{\lambda^{2}}{2\pi^2}K},
\ee
where $\epsilon'$ is a UV length scale, similar to but different from $\epsilon$, which may also depend on $\zeta$.

\begin{figure}[t!]
\begin{center}
\includegraphics[width=0.48\textwidth]{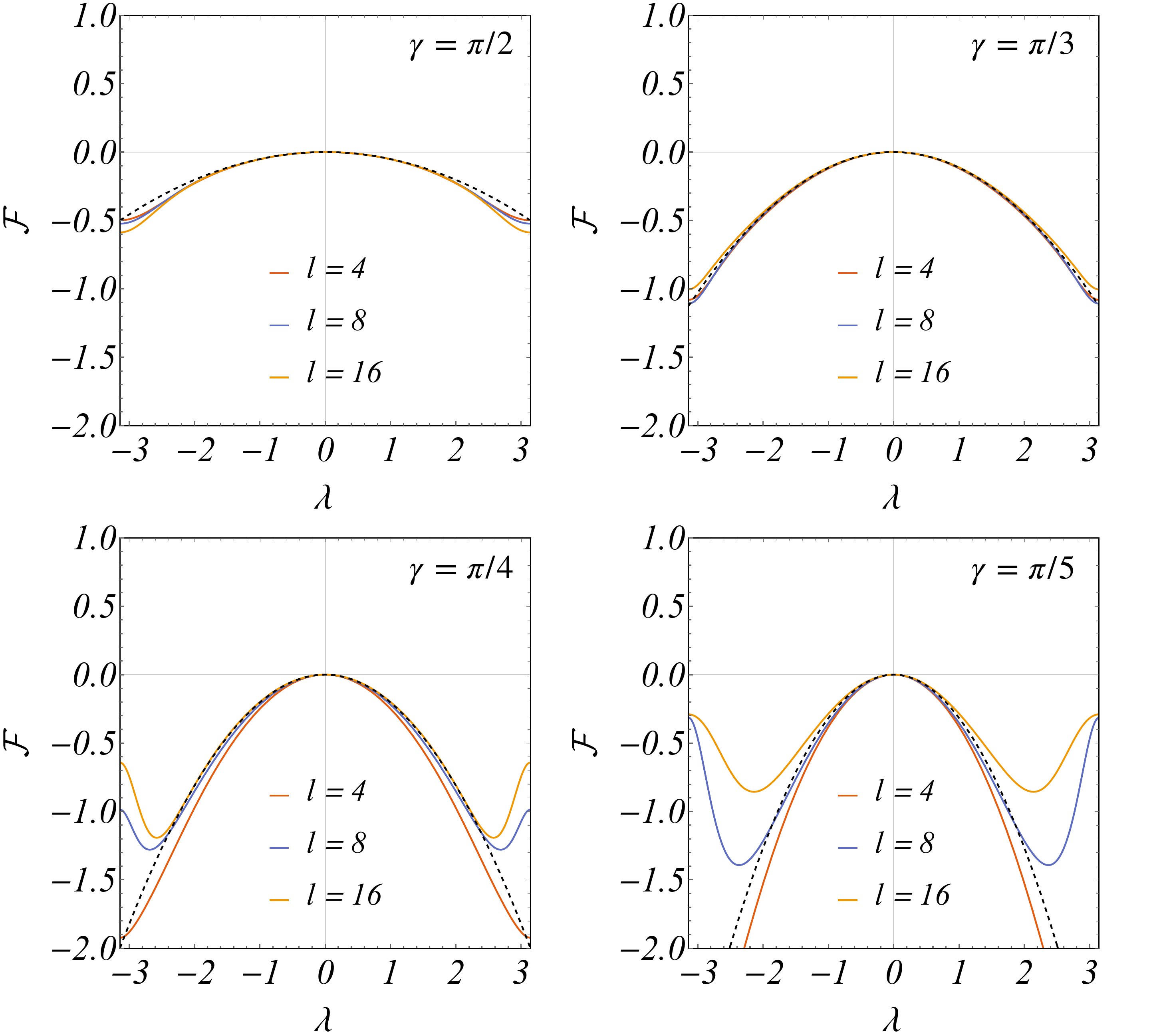}
\caption{\label{fig:F_ell}
The logarithm of the ratio of generating functions at the largest time $t \simeq 200$
for different subsystem sizes $l$ and parameter $\gamma$. 
The black dashed lines represent the LL quadratic prediction (\ref{eq:logratio}). 
}
\end{center}
\end{figure}

The numerical study of the full counting statistics in the LQSS is  
tricky due to its dependence on the subsystem size $l$. 
Indeed, Eq.~(\ref{eq:fullcountingscaling}) works only if the actual time reached by the unitary evolution is sufficiently large to guarantee 
a complete generalized thermalisation of the entire subsystem $[x-l/2,x+l/2]$. 
The dependence on $\epsilon'$ is canceled by considering the logarithm of the ratio between two different subsystem sizes; specifically,
\be\label{eq:logratio}
\mathcal{F}(\lambda) \equiv \log_2 \left[\frac{F_{[x-l,x+l]} ( \lambda, t)}{F_{[x-l/2,x+l/2]} ( \lambda, t)}\right] 
\underset{t \gg l \gg 1}{\simeq}
-\frac{K}{2\pi^2}\lambda^{2} .
\ee
Since both subsystems should be almost stationary, we focus on relatively small intervals, namely $l = 4, \,8$ and $16$.
In Fig.~\ref{fig:F_ell} we plot $\mathcal{F}(\lambda)$ for the subsystem at $x=0$ for $Q/P=1/2$, $1/3$, $1/4$ and $1/5$ at the maximum accessible time $t\simeq 200$.
Notice that the approach to the asymptotic behavior is not monotonic in $l$ (since
the information spreads out from the junction, the subsystem of size $2l$ takes 
longer to reach stationarity). 
Interestingly, all curves approach the stationary behavior from 
the neighborhood of $\lambda = 0$. For this reason, it is more instructive to
analyze the variance of the subsystem magnetization, i.e. the second cumulant,  
as a function of the subsystem size, both in the center of the system at $\zeta=0$ and away from it.

\begin{figure}[t!]
\begin{center}
\includegraphics[width=0.48\textwidth]{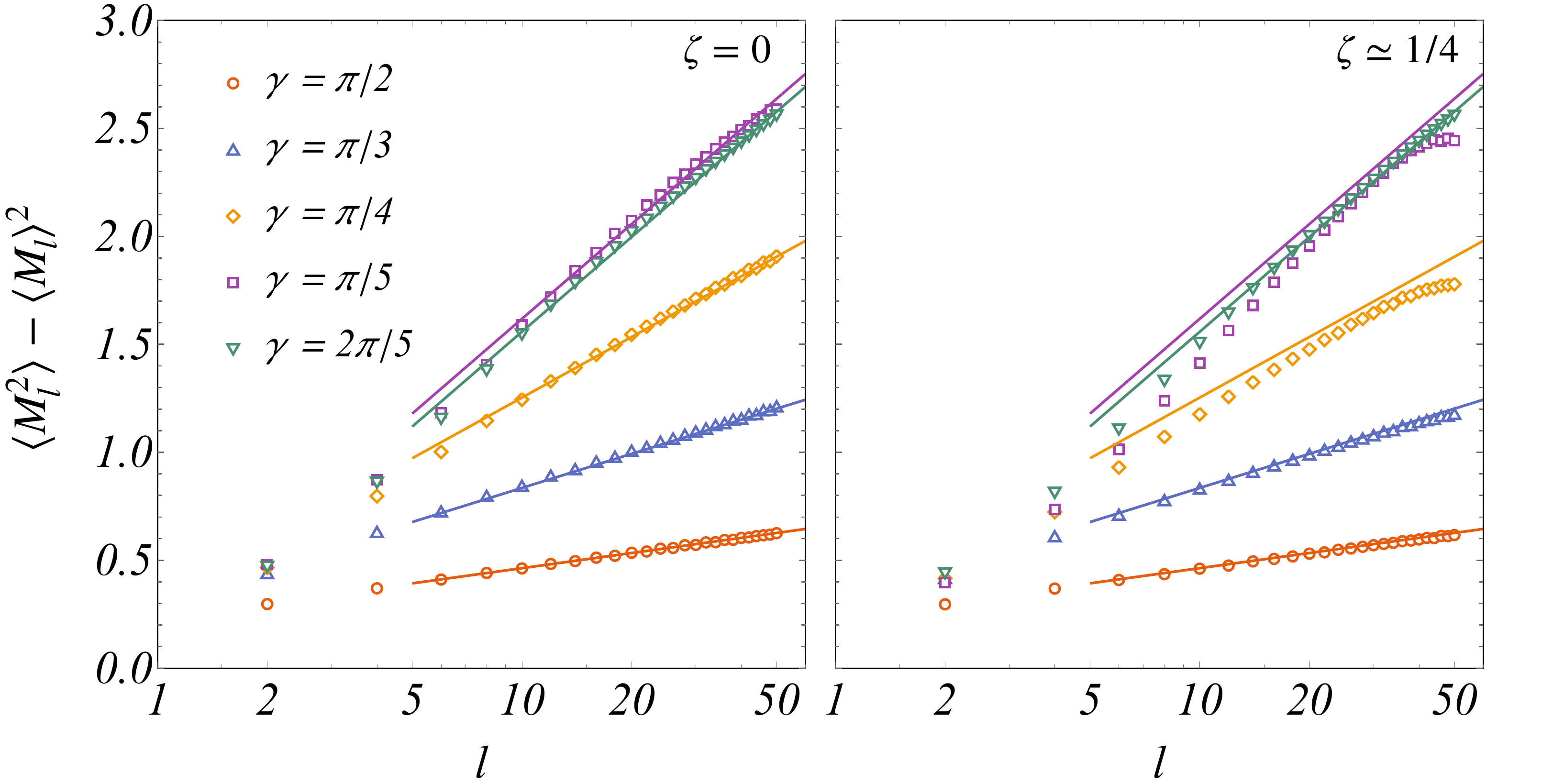} 
\caption{\label{fig:momenta}
The TEBD data for the variance of the subsystem magnetization (symbols) at the largest accessible time $t\simeq200$
are compared with the LL predictions (full lines). 
The subsystems are centered around $\zeta=0$ (left) and $\zeta=1/4$ (right).
}
\end{center}
\end{figure}

\paragraph{Fluctuations of subsystem magnetization. ---}The variance of the magnetization in the interval $[x - l/2, x+l/2]$ follows from Eq.~(\ref{eq:fullcountingscaling}) as 
\begin{multline}
 \langle M^{2}_{[x - l/2, x+l/2]}\rangle_{t} - \langle M_{[x - l/2, x+l/2]}\rangle_{t}^{2}    =\\  - \frac{1}{2} \partial_\lambda^2 \left.\log F_{[x - l/2, x+l/2]}\right|_{\lambda=0}
 \underset{t \gg l \gg 1}{\simeq}  \frac{K}{\pi^2} \log(l) + O(1). 
\end{multline}
In Fig. \ref{fig:momenta} we show the results obtained at $\zeta = x/t = 0$ and $\zeta \simeq 1/4$ for $l\in[2,50]$ and at $t\simeq 200$. For non-zero $\zeta$ the numerical analysis is slightly more difficult: 
the convergence is unavoidably poorer than at $\zeta=0$ because 
the approach to the LQSS requires more time as we move away from the junction.
Some comments are due: 
{\it (1)} the numerical data for all $\zeta$ manifest an asymptotic tendency toward the right
logarithmic behavior also with the same non-universal constant; the full lines are indeed the same in both panels;
 {\it (2)} 
for $\zeta \simeq 1/4$ the data show larger finite-size/finite-time effects because 
the subsystem is closer to the propagating front (the center of the subsystem is located at $x = 50 $ for $t = 200$); hence the subsystem is not relaxed for large $l$ when the numerical data deviate from the scaling prediction. 
{\it (3)} very remarkably, the numerical simulations, both for $\zeta=0$ and $\zeta\simeq1/4$,
show the same asymptotic behavior for two very different values of the anisotropy, namely 
$\Delta = \cos(\pi/5) \simeq 0.809 $ and $\Delta = \cos(2\pi/5) \simeq 0.309 $, 
confirming that the only parameter entering in the large-scale/large-time 
description of the local quasi-stationary state is the square of the quasi-particle charge (\ref{KP}), which depends only on $P$, the denominator of $\gamma/\pi$, see Eq. (\ref{eq:Delta}).

\paragraph{Discussions and conclusions. ---}

In this Letter we analytically showed that the stationary state resulting from the melting of a domain wall  in an XXZ chain is described at low energy (i.e. large times and 
distances) by an emergent Luttinger liquid with Luttinger parameter nowhere continuous in $\Delta$.
We corroborate this surprising prediction by accurate numerical tensor network simulations which strongly support our finding, manifested in the central charge of the underlying 
field theory being 1 (from  the measure of the entanglement entropy) and in the Luttinger parameter being $K=P^2/4$ (from measures of the magnetization statistics). 
It is remarkable that the Luttinger parameter has such fractal structure, because it implies that equal-time correlators have a fractal behavior.
This contrasts the nowadays well established results for spin Drude weight \cite{ID117,uoks-19,p-11,spa-09,z-99} which requires the measure of genuinely dynamical quantities.

In spite of these robust and intriguing findings, there are still many open questions. 
First, it would be interesting to determine correlation functions in the LQSS to provide further predictions to be tested numerically 
also to have further confirmations of the fractal Luttinger scenario; unfortunately, this is still beyond our technical capabilities. 
Another important question concerns the generality of our scenario:  
are there in more complicated integrable models (such as higher-spin chains or Hubbard models, studied already with GHD \cite{mbpc-18,nt-20,id-17}) 
zero entropy initial states with an LQSS being a fractal Luttinger liquid?  
What is the nature of the LQSS at the isotropic point $\Delta=1$ with pathological transport~\cite{mpp-19}?

Finally, we note that the domain-wall problem resembles the spatiotemporal quench protocol \cite{agarwal2018fast} for fast preparation of quantum critical systems. The idea is that, contrary to low-energy states of gapless systems ---which cannot be reached easily by cooling because temperatures would have to be prohibitively low---, product states can be engineered easily in cold atom experiments \cite{bernien2017probing}, and then be evolved unitarily. Thus, the domain-wall melting problem can be viewed as a realistic protocol for fast preparation of a Luttinger liquid, similarly to the protocol of Ref.~\cite{agarwal2018fast}. Our results show that the critical system engineered in this way will indeed be a Luttinger liquid, but it will be very different from the one corresponding to the ground state of the XXZ chain.

\paragraph{Acknowledgements. ---} We thank J. Viti for useful discussions and collaboration on related topics.
PC  acknowledges support from ERC under Consolidator grant  number 771536 (NEMO).

\onecolumngrid
\newpage 

\appendix
\setcounter{equation}{0}
\setcounter{figure}{0}
\renewcommand{\thetable}{S\arabic{table}}
\renewcommand{\theequation}{S\arabic{equation}}
\renewcommand{\thefigure}{S\arabic{figure}}

\begin{center}
{\Large Supplementary Material \\ 
\titleinfo
}
\end{center}
\section{Numerical methods.} 
The comparison between the analytical predictions from the Luttinger liquid description of the melted region and numerical results has been performed with the well-established TEBD algorithm~\cite{TEBD}. We perform numerical simulations for systems with $L=400$ lattice sites. The evolution operator is expanded using a $2^{\rm nd}$-order Suzuki-Trotter decomposition with time step $dt = 5\cdot 10^{-3}$. During the time evolution, the auxiliary dimension of the MPS representation of the state has been enlarged up to the maximum value $\chi_{\rm MAX}=512$. We were able to reach the maximum time $T\simeq 200$ by keeping the integrated truncation error below $\sim 10^{-8}$.

\section{Summary of integrable structure for XXZ at $|\Delta|<1$}
\subsection{Fine structure of quasiparticle content}
Here, we provide a short summary of the relevant functions involved in the thermodynamics and in the hydrodynamic description of the XXZ spin chain in the $|\Delta|<1$ region. We parameterize $\Delta = \cos(\gamma)$. We introduce the continued fraction representation
\begin{equation}
\frac{\gamma}{\pi}  = \frac{1}{\nu_1 + \frac{1}{\nu_2 + \ldots}} = [0; \nu_1, \nu_2, \ldots].
\end{equation}
As explained in the main text, we focus on rational points $\gamma/\pi = Q/P$, for which there is a finite set of $\nu$'s and $Q/P = [0; \nu_1,\ldots,\nu_\delta]$. The general case can then be recovered in the limit $\delta \to \infty$. 
Let us denote as $\mu_{i} = \sum_{j=1}^{i} \nu_{j}$, i.e. the partial sums of the $\nu$'s with $\mu_{0} = 0$; then,  the number of species is $\ell = \mu_\delta$. We introduce the approximants of $Q/P$ as the truncated continued fractions 
\begin{equation}
\frac{Q_k}{P_k} = [0; \nu_1,\ldots, \nu_k] \;, \qquad k =1,\ldots, \delta.
\end{equation}
Clearly, $Q_\delta = Q$ and $P_\delta = P$. Explicitly, one has the recursive relations
\begin{align}
P_{k} & = \nu_{k} P_{k-1} + P_{k-2}\; , \quad P_{0} = 1 \;,\quad P_{-1} = 0\;, \\
Q_{k} & = \nu_{k} Q_{k-1} + Q_{k-2}\; , \quad Q_{0} = 0 \;,\quad Q_{-1} = 1\;, \\
\end{align}
The difference between approximants satisfies
\begin{equation}
\label{approdiff}
 \frac{Q_k}{P_k} - \frac{Q_{k-1}}{P_{k-1}} = \frac{(-1)^{k+1}}{P_k P_{k-1}} \;.
\end{equation}
To each species $j = 1,\ldots, \ell$, it is associated a size $\nn_j \in \mathbb{N}$ and a parity $\pp_j \in \{0,1\}$, expressed as
\begin{align}
\nn_{j} &= P_{i-1} + (j-\mu_{i}) P_{i} \quad {\rm for} \quad \mu_{i} \le j < \mu_{i+1} \; \\
 \pp_{\mu_1} &= -1 \;,  \quad \pp_{j} = (-1)^{\left\lfloor (\nn_{j} - 1)\frac{Q}{P} \right\rfloor} \quad  {\rm for} \quad j \neq \mu_{1} \; .
\end{align}
Finally, let us collect some useful relations involving the last strings
(where we used the definition $\ell = \mu_{\delta}$ and the fact that $P_{\delta} = P$):
\begin{align}
\label{nident}
\nn_{\ell} = P_{\delta - 1}\; \quad \nn_{\ell} + \nn_{\ell-1} = P \; , \quad  \nn_{\ell-1} - \nn_{\ell} = \nn_{\ell-2}. 
\end{align}
As mentioned in the main text, an equilibrium state is specified by a set of filling functions $\{\vartheta_j(\lambda)\}_{j=1}^{\ell}$, with $0\leq \vartheta_j(\lambda) \leq 1$. 

\subsection{Scattering kernel and dressing}
The interaction between the $\ell$ species defined above is encoded in the scattering kernel, which takes the form
\begin{align}
\label{Tdef}
T_{j,k}(\lambda) = (1- \delta_{\nn_j, \nn_k}) a_{|\nn_j - \nn_k|}^{(\pp_j \pp_k)}(\lambda) 
+ 2 a_{|\nn_j - \nn_k|+2}^{(\pp_j \pp_k)}(\lambda) 
+ \ldots + 2 a_{\nn_j + \nn_k - 2}^{(\pp_j \pp_k)}(\lambda)  + a_{\nn_j + \nn_k}^{(\pp_j \pp_k)}(\lambda),
\end{align}
where we introduced the function 
\be
a_{\nn}^{(\pp)}(\lambda) = \frac{\pp}{\pi} \frac{\sin(\gamma \nn)}{\cosh(2\lambda) - \pp \cos(\gamma \nn)} \; .
\ee
For instance, excitations on top of an equilibrium state specified by $\{\vartheta_j(\lambda)\}_{j=1}^{\ell}$ are described the dressed single-particle eigenvalues $q_j(\lambda)$, which can be obtained from the bare ones $\mathfrak{q}_j(\lambda)$ solving the dressing equation
\be\label{eq:dress_q}
q_j(\lambda) = \mathfrak{q}_j(\lambda) 
- \sum_{k=1}^\ell \int d\mu \,T_{j,k}(\lambda - \mu) \sigma_k \vartheta_k(\mu) q_k(\mu)\; .
\ee

\section{Summary of GHD solution from Domain-Wall initial state}
We report here the profile of the magnetization and spin current obtained via the GHD solution starting from the domain wall initial state. From the solution of the GHD equation \eqref{LQSSDW} in the main text, one obtains for a fixed $\zeta = x/t$
\begin{equation}
\label{eq:profiles}
\langle {\boldsymbol s}^{z}\rangle_{\zeta} 
 = - \frac{P}{2\pi} \arcsin \left( \frac{\zeta}{ \zeta_0}  \right)\; ,\qquad
\langle {\boldsymbol j}_{{\boldsymbol s}^{z}}\rangle_{\zeta} 
 = \frac{P}{2\pi} \zeta_0 
 \left [ \sqrt{1-  \frac {\zeta^2} {\zeta^2_0}} 
- \cos \left(\frac{\pi}{P} \right) \right] \; ,
\end{equation}
where $\zeta_0 = \sin(\gamma)/\sin(\pi/P)$.
\section{Calculation of the Luttinger parameter}
In this section, we compute the dressed magnetization at the Fermi points, for each ray in the LQSS emerging from the domain wall. 
Specifying Eq.~\eqref{eq:dress_q}, for the magnetization $n_j(\lambda)$ we arrive at
\be\label{eq:dress_m}
n_j(\lambda) = \nn_j 
- \sum_{k=1}^\ell \int d\mu \,T_{j,k}(\lambda - \mu) \sigma_k \vartheta_k(\mu) n_k(\mu)\; .
\ee
Then, from \eqref{Tdef} one finds that 
\begin{align}
& T_{\ell, \ell} (\lambda) =  T_{\ell-1, \ell-1} (\lambda)  = - T_{\ell - 1, \ell} (\lambda) =  -T_{\ell, \ell-1} (\lambda) \equiv \tau(\lambda) ,
\\ & \sigma_{\ell -1} = - \sigma_\ell. 
\end{align}
Moreover, for a fixed ratio $\zeta = x/t$, the filling functions $\vartheta_j(\lambda; \zeta = x/t)$ are given in \eqref{LQSSDW} and thus vanish for $k = 1,\ldots, \ell-2$; this allows us to restrict the sums in \eqref{eq:dress_m} to $k = \ell-1$ and $k = 
\ell$. Moreover, 
\begin{equation}
\vartheta_\ell(\lambda; \zeta) = \vartheta_{\ell-1}(\lambda; \zeta) = \begin{cases} 
                                                                   1 & \lambda > \lambda_\zeta^\ast \\
                                                                   0 & \mbox{otherwise}
                                                                  \end{cases} ,
\end{equation}
with the rapidity $\lambda_\zeta^\ast$ at the Fermi point is defined via 
\begin{equation}
v^{\rm eff}_\ell (\lambda_\zeta^\ast) = \zeta\;. 
\end{equation}
Therefore, we can rewrite \eqref{eq:dress_m} as
\begin{align}
n_{\ell-1}(\lambda) &= \nn_{\ell-1} 
+ \sigma_\ell \int_{\lambda_\zeta^\ast}^\infty d\mu \,\tau(\lambda - \mu)  (n_{\ell-1}(\mu) + n_\ell(\mu))\; ,\\
n_{\ell}(\lambda) &= \nn_{\ell} 
- \sigma_\ell \int_{\lambda_\zeta^\ast}^\infty d\mu \,\tau(\lambda - \mu)  (n_{\ell-1}(\mu) + n_\ell(\mu))\; .\end{align}
Using \eqref{nident}, we get 
\begin{equation}
 n_{\ell-1}(\lambda) + n_{\ell}(\lambda) = P,
\end{equation}
and consequently
\begin{equation}
n_{\ell-1}(\lambda) - n_{\ell}(\lambda) = \nn_{\ell-2} 
+ 2 \sigma_\ell P \int_{\lambda_\zeta^\ast}^\infty d\mu \,\tau(\lambda - \mu) .
\end{equation}
For $\lambda = \lambda_\zeta^\ast$, using $\tau(\lambda) = \tau(-\lambda)$, we obtain
\begin{equation}
n_{\ell-1}(\lambda_\zeta^\ast) - n_{\ell}(\lambda_\zeta^\ast) = \nn_{\ell-2} 
+ \sigma_\ell P \int_{-\infty}^\infty d\mu \,\tau(\mu)  \;.
\end{equation}
The right-hand side can be further simplified using that
\begin{equation}
\label{aint}
 \int_{-\infty}^\infty d\lambda \; a_n^{(1)}(\lambda) = 
 1 - 2 \left\{ \frac{n \gamma}{2\pi}\right\},
\end{equation}
where $\{x\}$ stands for the fractional part of the real number $x$. Indeed, from \eqref{aint} and \eqref{Tdef}
one has
\begin{equation}
  \int_{-\infty}^\infty d\lambda \; \tau(\lambda) = \frac{1 - \sigma_\ell}{2} - \left\{ \frac{2 P_{\delta -1}^2 \gamma}{\pi} \right\} = - \sigma_\ell \left[1 - \left\{ \frac{2 P_{\delta -1}}{P} \right\} \right],
\end{equation}
where in the last equality we used \eqref{approdiff} for $k = \delta$. Finally, from \eqref{nident}, we obtain that $\nn_{\ell-2} = P - 2 P_{\delta -1}$, which implies 
\begin{equation}
 n_{\ell-1}(\lambda_\zeta^\ast) = n_{\ell}(\lambda_\zeta^\ast) =  P/2,
\end{equation}
as we claimed. 


\begin{thebibliography}{99}

\bibitem{torrance1969excitation}
J. Torrance and M. Tinkham,  \textit{Excitation of Multiple-Magnon Bound States in ${\rm CoCl}_2 \cdot 2{\rm H}_2 {\rm O}$},  
\href{\doi10.1103/PhysRev.187.595}{Phys. Rev. {\bf 187}, 595 (1969)}.

\bibitem{gochev1977spin}
I.G. Gochev,   {\it Spin complexes in a bounded chain},  
\href{http://www.jetpletters.ac.ru/ps/1375/article_20820.pdf}{JETP {\bf 26}, 3 (1977)}.

\bibitem{gochev1983contribution}
I.G. Gochev,  {\it Contribution to the theory of plane domain walls in a ferromagnet}, 
\href{http://www.jetp.ac.ru/cgi-bin/dn/e_058_01_0115.pdf}{JETP {\bf 58}, 115 (1983)}.

\bibitem{yuan2007domain}
 S. Yuan, H. De Raedt, and S. Miyashita,  {\it Domain-wall dynamics near a quantum critical point},  
 \href{https://doi.org/10.1103/PhysRevB.75.184305}{Phys. Rev. B {\bf 75}, 184305 (2007).}

\bibitem{antal1999transport}
  T. Antal, Z. R{\'a}cz, A. R{\'a}kos, and G. Sch{\"u}tz,  {\it Transport in the XX chain at zero temperature: Emergence of flat magnetization profiles},  
  \href{https://doi.org/10.1103/PhysRevE.59.4912}{Phys. Rev. E {\bf 59}, 4912 (1999).}

\bibitem{hunyadi2004dynamic}
 V. Hunyadi, Z. R{\'a}cz, and L. Sasv{\'a}ri,  {\it Dynamic scaling of fronts in the quantum XX chain},
 \href{https://doi.org/10.1103/PhysRevE.69.066103}{Phys. Rev. E {\bf 69}, 066103 (2004).}

\bibitem{platini2005scaling}
T. Platini and D. Karevski,  {\it Scaling and front dynamics in Ising quantum chains},  \href{https://doi.org/10.1140/epjb/e2005-00402-2}{Europ. Phys. J. B {\bf 48}, 225 (2005).}

\bibitem{platini2007relaxation}
T. Platini and D. Karevski,  {\it Relaxation in the XX quantum chain},  \href{https://doi.org/10.1088/1751-8113/40/8/002}{J. Phys. A {\bf 40}, 1711 (2007).}

\bibitem{NESSf5}
A. De Luca, J. Viti, D. Bernard, and B. Doyon, 
\textit{Nonequilibrium thermal transport in the quantum Ising chain},
 \href{\doi10.1103/PhysRevB.88.134301}{Phys. Rev. B {\bf 88}, 1342301 (2013)}.

\bibitem{NESSf6}
A. De Luca, G. Martelloni, and J. Viti,  \textit{Stationary states in a free fermionic chain from the quench action method},
\href{\doi10.1103/PhysRevA.91.021603}{Phys. Rev. A {\bf 91}, 021603 (2014)}. 

\bibitem{viti2016inhomogeneous}
  J. Viti, J.-M. St{\'e}phan, J. Dubail, and M. Haque,  
 {\it Inhomogeneous quenches in a free fermionic chain: Exact results}, 
 \href{http://doi.org/10.1209/0295-5075/115/40011}{Europhysics Lett. {\bf 115}, 40011 (2016).}  

\bibitem{gobert2005real}
 D. Gobert, C. Kollath, U. Schollw\"ock, and G. M. Sch\"utz,  
 {\it Real-time dynamics in spin-1/2 chains with adaptive time-dependent density matrix renormalization group},  
 \href{https://doi.org/10.1103/PhysRevE.71.036102}{Phys. Rev. E {\bf 71}, 036102 (2005).}

\bibitem{jesenko2011finite}
S. Jesenko and M. Znidaric,  {\it Finite-temperature magnetization transport of the one-dimensional anisotropic Heisenberg model},  
\href{https://doi.org/PhysRevB.84.174438}{Phys. Rev. B {\bf 84}, 174438 (2011)}.

\bibitem{chl-08}
P. Calabrese, C. Hagendorf, and P. Le Doussal,
{\it Time evolution of 1D gapless models from a domain-wall initial state: SLE continued?},
\href{https://doi.org/10.1088/1742-5468/2008/07/P07013}{J. Stat. Mech. P07013 (2008)}.

\bibitem{zauner2012time}
  V. Zauner, M. Ganahl, H. Evertz, and T. Nishino,  {\it Time Evolution within a Comoving Window: Scaling of signal fronts and magnetization plateaus after a local quench in quantum spin chains},  \href{https://doi.org/10.1088/0953-8984/27/42/425602}{J. Phys.: Condens. Matter {\bf 27}, 425602 (2012).}

\bibitem{halimeh2014domain}
J. Halimeh, A. W\"ollert, I. McCulloch, U. Schollw\"ock, and T. Barthel,  
{\it Domain-wall melting in ultracold-boson systems with hole and spin-flip defects}, 
\href{https://doi.org/10.1103/PhysRevA.89.063603}{Phys. Rev. A {\bf 89}, 063603 (2014).}

\bibitem{alba2014entanglement}
V. Alba and F. Heidrich-Meisner,  {\it Entanglement spreading after a geometric quench in quantum spin chains}, 
\href{https://doi.org/10.1103/PhysRevB.90.075144}{Phys. Rev. B {\bf 90}, 075144 (2014).}

\bibitem{hauschild2016domain}
J. Hauschild, F. Heidrich-Meisner, and F. Pollmann,  {\it Domain-wall melting as a probe of many-body localization}, 
 \href{https://doi.org/10.1103/PhysRevB.94.161109}{Phys. Rev. B {\bf 94}, 161109(R) (2016).}

\bibitem{NESSnum1}
 T. Sabetta and G. Misguich, \textit{Nonequilibrium steady states in the quantum XXZ spin chain},
 \href{\doi10.1103/PhysRevA.91.021603}{Phys. Rev. B {\bf 88}, 245114 (2013)}.    

\bibitem{NESSnum2}
 C. Karrasch, R. Ilan, and J. E. Moore, \textit{Nonequilibrium thermal transport and its relation to linear response},
 \href{https://doi.org/10.1103/PhysRevB.88.195129}{Phys. Rev. B {\bf  88}, 195129 (2013)}.

\bibitem{NESSnum3}
 A. Biella, A. De Luca, J. Viti, D. Rossini, L. Mazza, and R. Fazio,
\textit{Energy transport between two integrable spin chains},
\href{https://doi.org/10.1103/PhysRevB.93.205121}{Phys. Rev. B {\bf 93}, 205121 (2016)}.

\bibitem{DVMR14}
 A. De Luca, J. Viti, L. Mazza, and D. Rossini, \textit{Energy transport in Heisenberg chains beyond the Luttinger liquid paradigm}, 
\href{http://doi.org/10.1103/PhysRevB.90.161101}{Phys. Rev. B {\bf 90}, 161101 (2014)}.

\bibitem{bernarddoyon2016}
D. Bernard and B. Doyon, 
\textit{Conformal field theory out of equilibrium: a review},
\href{\doi10.1088/1742-5468/2016/06/064005}{J. Stat. Mech. {064005} (2016)}.

\bibitem{NESSf7}
 B. Doyon, A. Lucas, K. Schalm, and M. J. Bhaseen, \textit{Non-equilibrium steady states in the Klein-Gordon theory},
 \href{https://doi.org/10.1088/1751-8113/48/9/095002}{J. Phys. A {\bf 48}, 095002 (2015)}.

\bibitem{NESSf11}
 L. Vidmar, D. Iyer, and M. Rigol, \textit{Emergent Eigenstate Solution to Quantum Dynamics Far from Equilibrium},
 \href{\doi10.1103/PhysRevX.7.021012}{Phys. Rev. X {\bf 7}, 021012 (2017)}.

\bibitem{rakovszky2019entanglement}
T. Rakovszky, C. von Keyserlingk, and F. Pollmann,  {\it Entanglement growth after inhomogenous quenches},  
  \href{https://doi.org/10.1103/PhysRevB.100.125139}{Phys. Rev. B {\bf 100}, 125139 (2019).}

\bibitem{bulchandani2019subdiffusive}
V. Bulchandani and C. Karrasch, {\it Subdiffusive front scaling in interacting integrable models}, \href{https://doi.org/10.1103/PhysRevB.99.121410}{Phys. Rev. B {\bf 99}, 121410 (2019).}

\bibitem{bertini2016transport}
B. Bertini, M. Collura, J. De Nardis, and M. Fagotti,  {\it Transport in Out-of-Equilibrium XXZ Chains: Exact Profiles of Charges and Currents},  
\href{https://doi.org/10.1103/PhysRevLett.117.207201}{Phys. Rev. Lett. {\bf 117}, 207201 (2016).}

\bibitem{castro2016emergent}
 O. Castro-Alvaredo, B. Doyon, and T. Yoshimura,   {\it Emergent Hydrodynamics in Integrable Quantum Systems Out of Equilibrium},   
 \href{https://doi.org/10.1103/PhysRevX.6.041065}{Phys. Rev. X {\bf 6}, 041065 (2016)}. 

\bibitem{doytaka2017}
B. Doyon, and T. Yoshimura, \textit{A note on generalized hydrodynamics: inhomogeneous fields and other concepts},
\href{\doi10.21468/SciPostPhys.2.2.014}{SciPost Phys. {\bf 2}, 014 (2017)}. 

\bibitem{DDKY17}
B. Doyon, J. Dubail, R. Konik, and T. Yoshimura,  
\textit{Large-Scale Description of Interacting One-Dimensional Bose Gases: Generalized Hydrodynamics Supersedes Conventional Hydrodynamics},
\href{https://doi.org/10.1103/PhysRevLett.119.195301}{Phys. Rev. Lett. {\bf 119}, 195301 (2017)}.

\bibitem{bulchandani2018bethe}
V. Bulchandani, R. Vasseur, C. Karrasch, and J. Moore, {\it Bethe-Boltzmann Hydrodynamics and Spin Transport in the XXZ Chain}, \href{https://doi.org/10.1103/PhysRevB.97.045407}{Phys. Rev. B {\bf 97}, 045407 (2018).}

\bibitem{bastianelloalba2019}
A. Bastianello, V. Alba, and J.-S. Caux, 
\textit{Generalized hydrodynamics with space-time inhomogeneous interactions},
\href{\doi10.1103/PhysRevLett.123.130602}{Phys. Rev. Lett. {\bf123}, 130602 (2019)}. 

\bibitem{ms-20}
F. S. Moller and J. Schmiedmayer, {\it Introducing iFluid: a numerical framework for solving hydrodynamical equations in integrable models},
\href{https://arxiv.org/pdf/2001.02547.pdf}{ArXiv:2001.02547}.

\bibitem{BDWY17}
A. Bastianello, B. Doyon, G. Watts, and T. Yoshimura,
\textit{Generalized hydrodynamics of classical integrable field theory: the sinh-Gordon model},
\href{\doi10.21468/SciPostPhys.4.6.045}{SciPost Phys. {\bf 4}, 045 (2018)}.

\bibitem{ds-17}
B. Doyon and H. Spohn, {\it Dynamics of hard rods with initial domain wall state},
\href{https://doi.org/10.1088/1742-5468/aa7abf}{J. Stat. Mech. 2017, 073210 (2017)}.

\bibitem{dsy-17}
B. Doyon, H. Spohn, and T. Yoshimura, {\it A geometric viewpoint on generalized hydrodynamics},
\href{https://doi.org/10.1016/j.nuclphysb.2017.12.002}{Nucl. Phys. B {\bf 926}, 570 (2017)}

\bibitem{bib:cdv18}
M. Collura, A. De Luca, and J. Viti, 
{\it Analytic solution of the domain-wall nonequilibrium stationary state},
\href{https://doi.org/10.1103/PhysRevB.97.081111}{Phys. Rev. B {\bf 97}, 081111(R) (2018).}

\bibitem{ghddiff}
J. De Nardis, D. Bernard, and B. Doyon, \textit{Hydrodynamic diffusion in integrable systems},
\href{https://doi.org/10.1103/PhysRevLett.121.160603}{Phys. Rev. Lett. {\bf 121}, 160603 (2018)}.

\bibitem{vasseurdiff}
S. Gopalakrishnan, D. A. Huse, V. Khemani, and R. Vasseur, 
\textit{Hydrodynamics of operator spreading and quasiparticle diffusion in interacting integrable systems},
\href{https://doi.org/10.1103/PhysRevB.98.220303}{Phys. Rev. B {\bf 98}, 220303 (2018)}.

\bibitem{milosz2019}
M. Panfil and J. Pawe\l czyk,
{\it Linearized regime of the generalized hydrodynamics with diffusion},
\href{https://arxiv.org/abs/1905.06257}{arXiv:1905.06257}.

\bibitem{superdiffusive2018}
E. Ilievski, J. De Nardis, M. Medenjak, and T. Prosen, \textit{Superdiffusion in one-dimensional quantum lattice models},
\href{https://doi.org/10.1103/PhysRevLett.121.230602}{Phys. Rev. Lett. {\bf 121}, 230602 (2018)}.

\bibitem{doyonsuper}
B. Doyon, \textit{Diffusion and superdiffusion from hydrodynamic projection}, \href{https://arxiv.org/abs/1912.01551}{arXiv:1912.01551} (2019).




\bibitem{nonball2017}
L. Piroli, J. De Nardis, M. Collura, B. Bertini, and M. Fagotti,\textit{ Transport in out-of-equilibrium XXZ chains: Nonballistic behavior and correlation functions},
\href{\doi10.1103/PhysRevB.96.115124}{Phys. Rev. B {\bf 96}, 115124 (2017)}

\bibitem{kapitza2019}
A. Biella, M. Collura, D. Rossini, A. De Luca, and L. Mazza,
\textit{Ballistic transport and boundary resistances in inhomogeneous quantum spin chains},
 \href{\doi10.1038/s41467-019-12784-4}{Nature Commun. {\bf 10}, 4820 (2019)}.

\bibitem{friedman2019}
A. J. Friedman, S. Gopalakrishnan, and R. Vasseur,
\textit{Diffusive hydrodynamics from integrability breaking},
\href{https://arxiv.org/abs/1912.08826}{arXiv: 1912.08826} (2019).

\bibitem{flux2019}
A. Bastianello and A. De Luca, \textit{Integrability-protected adiabatic reversibility in quantum spin chains}, 
\href{\doi10.1103/PhysRevLett.122.240606}{Phys. Rev. Lett. {\bf 122}, 240606 (2019)}.

\bibitem{quasilongrange}
J. De Nardis and M. Panfil, 
\textit{Edge Singularities and Quasilong-Range Order in Nonequilibrium Steady States},
\href{https://doi.org/10.1103/PhysRevLett.120.217206}{Phys. Rev. Lett. {\bf 120}, 217206 (2018)}.

\bibitem{fagotti2017}
M. Fagotti, \textit{Higher-order generalized hydrodynamics in one dimension: The noninteracting test},
\href{\doi10.1103/PhysRevB.96.220302}{Phys. Rev. B {\bf 96}, 220302 (2017)}.

\bibitem{fagotti2019}
M. Fagotti, \textit{Locally quasi-stationary states in noninteracting spin chains},
\href{https://arxiv.org/abs/1910.01046}{arXiv:1910.01046} (2019).

\bibitem{brun2018}
Y. Brun and J. Dubail, \textit{The Inhomogeneous Gaussian Free Field, with application to ground state correlations of trapped 1d Bose gases}, 
\href{\doi10.21468/SciPostPhys.4.6.037}{SciPost Phys. {\bf 4}, 037 (2018)}.

\bibitem{cftbreathing2019}
P. Ruggiero, Y. Brun, and J. Dubail, 
\textit{Conformal field theory on top of a breathing one-dimensional gas of hard core bosons},
\href{\doi10.21468/SciPostPhys.6.4.051}{SciPost Phys. {\bf 6}, 051 (2019)}. 

\bibitem{BFPC18}
B. Bertini, M. Fagotti, L. Piroli, and P. Calabrese, 
\textit{Entanglement evolution and generalised hydrodynamics: noninteracting systems},
\href{\doi10.1088/1751-8121/aad82e}{J. Phys. A {\bf 51}, 39LT01 (2018)}.

\bibitem{alba2019}
V. Alba, {\it Towards a generalized hydrodynamics description of Rényi entropies in integrable systems},
\href{https://doi.org/10.1103/PhysRevB.99.045150}{Phys. Rev. B {\bf 99}, 045150 (2019)}.

\bibitem{BAF19}
V. Alba, B. Bertini, and M. Fagotti, 
\textit{Entanglement evolution and generalised hydrodynamics: Interacting integrable systems},
\href{https://scipost.org/SciPostPhys.7.1.005}{SciPost Phys. {\bf 7}, 005 (2019)}.

\bibitem{bastiadubail}
A. Bastianello, J. Dubail, and J.-M. St\'ephan, 
\textit{Entanglement entropies of inhomogeneous Luttinger liquids},
\href{https://arxiv.org/abs/1910.09967}{arXiv:1910.09967}.

\bibitem{fokkema2014split}
	T. Fokkema, S. Eli\"ens, and J.-S. Caux, {\it Split Fermi seas in one-dimensional Bose fluids}, \href{https://doi.org/10.1103/PhysRevA.89.033637}{Phys. Rev. A {\bf 89}, 033637 (2014)}.

\bibitem{eliensjs}
S. Eli\"ens, and J.-S. Caux,\textit{General finite-size effects for zero-entropy states in one-dimensional quantum integrable models}, \href{https://doi.org/10.1088/1751-8113/49/49/495203}{J. Phys. A {\bf 49}, 495203 (2016)}.

\bibitem{correlations2016vlijm}
	R. Vlijm, S. Eli\"ens, J. -S. Caux, {\it Correlations of zero-entropy critical states in the XXZ model: integrability and Luttinger theory far from the ground state}, \href{https://doi.org/10.21468/SciPostPhys.1.1.008}{SciPost Phys. {\bf 1}, 008 (2016)}.

\bibitem{ruggiero2019}
P. Ruggiero, P. Calabrese, B. Doyon, and J. Dubail, 
\textit{Quantum Generalized Hydrodynamics},
\href{https://arxiv.org/abs/1910.00570}{arXiv:1910.00570}      



\bibitem{bertinifagotti2016}
B. Bertini and M. Fagotti,
\textit{Determination of the Nonequilibrium Steady State Emerging from a Defect},
\href{\doi10.1103/PhysRevLett.117.130402}{Phys. Rev. Lett. {\bf 117}, 130402 (2016)}.

\bibitem{Bas_Deluca_defhop}
 A. Bastianello and A. De Luca, 
 \textit{Nonequilibrium Steady State Generated by a Moving Defect: The Supersonic Threshold},
 \href{https://doi.org/10.1103/PhysRevLett.120.060602}{Phys. Rev. Lett.  {\bf 120}, 060602 (2018)}.

\bibitem{Bas_Deluca_defising}
 A. Bastianello and A. De Luca,  
 \textit{Superluminal moving defects in the Ising spin chain},
 \href{https://doi.org/10.1103/PhysRevB.98.064304}{Phys. Rev. B {\bf 98}, 064304 (2018)}.

\bibitem{giamarchi2004one}
  T. Giamarchi, {\it Quantum physics in one dimension}, Clarendon press (2003).

\bibitem{sirker2005}
J. Sirker and M. Bortz, 
\textit{The open XXZ-chain: Bosonisation, Bethe ansatz and logarithmic corrections},
\href{\doi10.1088/1742-5468/2006/01/P01007}{J. Stat. Mech. {P01007} (2006)}. 

\bibitem{bpc-18}
B. Bertini, L. Piroli, and P. Calabrese, {\it Universal broadening of the light cone in low-temperature transport}, 
\href{\doi10.1103/PhysRevLett.120.176801}{Phys. Rev. Lett. {\bf 120}, 176801 (2018)}.

\bibitem{bp-18}
B. Bertini and L. Piroli, {\it Low-Temperature Transport in Out-of-Equilibrium XXZ Chains},
\href{\doi10.1088/1742-5468/aab04b}{J. Stat. Mech. (2018) 033104}.

\bibitem{PrIl13}
T. Prosen and E. Ilievski, 
\textit{Families of Quasilocal Conservation Laws and Quantum Spin Transport}, 
\href{http://dx.doi.org/10.1103/PhysRevLett.111.057203}{Phys. Rev. Lett. {\bf 111}, 57203 (2013)}.

\bibitem{ID117}
E. Ilievski and J. De Nardis, 
\textit{Microscopic Origin of Ideal Conductivity in Integrable Quantum Models},
\href{http://dx.doi.org/10.1103/PhysRevLett.119.020602}{Phys. Rev. Lett.  {\bf 119}, 020602 (2017)}.

\bibitem{bib:mmk17}
G. Misguich, K. Mallick, and P. L. Krapivsky,
{\it Dynamics of the spin-$1/2$ Heisenberg chain initialized in a domain-wall state},
\href{https://doi.org/10.1103/PhysRevB.96.195151}{Phys. Rev. B {\bf 96}, 195151 (2017).}

\bibitem{LZP17}
M. Ljubotina, M. Znidaric, and T. Prosen, 
\textit{Spin diffusion from an inhomogeneous quench in an integrable system},
\href{http://dx.doi.org/10.1038/ncomms16117}{Nature Commun. {\bf 8}, 16117 (2017)}.

\bibitem{mpp-19}
G. Misguich, N. Pavloff, and V. Pasquier, {\it Domain wall problem in the quantum XXZ chain and semiclassical behavior close to the isotropic point},
\href{http://dx.doi.org/10.21468/SciPostPhys.7.2.025}{SciPost Phys. {\bf 7}, 025 (2019)}.

\bibitem{takahashi}
 M. Takahashi, {\it Thermodynamics of one-dimensional solvable models}, Cambridge University Press (1999).  

\bibitem{Note1}
See supplementary material at [url] for details about: numerical implementation; a summary of the GHD solution of the DW state; the calculation of the Luttinger parameter from the dressed magnetizations.

\bibitem{riemann1860}
B. Riemann,  {\it \"Uber die Fortpflanzung ebener Luftwellen von endlicher Schwingungsweite},  Abhandlungen der Geselschaft der Wissenschaften zu G\"ottingen, Mathematisch-Physicalische Klasse 8, 43 (1860).  

\bibitem{dubail2017conformal}
J. Dubail, J.-M. St{\'e}phan, J. Viti, and P. Calabrese,  
{\it Conformal field theory for inhomogeneous one-dimensional quantum systems: the example of non-interacting Fermi gases},
\href{https://doi.org/10.21468/SciPostPhys.2.1.002}{SciPost Phys. {\bf 2}, 002 (2017).}

\bibitem{le2008charge}
 K. Le Hur, B. Halperin, and A. Yacoby, {\it Charge fractionalization in nonchiral Luttinger systems},
 \href{https://doi.org/10.1016/j.aop.2008.04.006}{Ann. Phys. {\bf 323}, 3037 (2008)}.

\bibitem{dubail2017emergence}
 J. Dubail, J.-M. St\'ephan and P. Calabrese, {\it Emergence of curved light-cones in a class of inhomogeneous Luttinger liquids}, \href{http://dx.doi.org/10.21468/SciPostPhys.3.3.019}{SciPost Phys. {\bf 3}, 019 (2017)}.

\bibitem{allegra2016inhomogeneous}
 N. Allegra, J. Dubail, J.-M. St{\'e}phan and J. Viti,  {\it Inhomogeneous field theory inside the arctic circle}, 
  \href{https://doi.org/10.1088/1742-5468/2016/05/053108}{J. Stat. Mech. 053108 (2016).}

\bibitem{granet2019inhomogeneous}
 E. Granet, L. Budzynski, J. Dubail, and J. Jacobsen,  {\it Inhomogeneous Gaussian free field inside the interacting arctic curve},  \href{https://doi.org/10.1088/1742-5468/aaf71b}{J. Stat. Mech. 013102 (2019).}

\bibitem{calacardy2009}
P. Calabrese and J. Cardy, \textit{Entanglement entropy and conformal field theory},
\href{\doi10.1088/1751-8113/42/50/504005}{J. Phys. A {\bf 42}, 504005 (2009)}.

\bibitem{TEBD}
G. Vidal, \textit{Efficient simulation of one-dimensional quantum many-body systems},
\href{	\doi10.1103/PhysRevLett.93.040502}{Phys. Rev. Lett. {\bf 93}, 040502 (2004)}.

\bibitem{z-99}
X. Zotos, {\it Finite temperature drude weight of the one-dimensional spin-1/2 Heisenberg model}, 
\href{	\doi10.1103/PhysRevLett.82.1764}{Phys. Rev. Lett. {\bf 82}, 1764 (1999)}.

\bibitem{spa-09}
J. Sirker, R. G. Pereira, and I. Affleck, {\it Diffusion and ballistic transport in one-dimensional quantum systems}, 
\href{	\doi10.1103/PhysRevLett.103.216602}{Phys. Rev. Lett. {\bf 103}, 216602 (2009)}.

\bibitem{p-11}
T. Prosen, {\it Open xxz spin chain: Nonequilibrium steady state and a strict bound on ballistic transport}, 
\href{	\doi10.1103/PhysRevLett.106.217206}{Phys. Rev. Lett. {\bf 106}, 217206 (2011)}.

\bibitem{uoks-19}
A. Urichuk, Y. Oez, A. Klumper, J. Sirker, {\it The spin Drude weight of the XXZ chain and generalized hydrodynamics},
\href{	\doi10.21468/SciPostPhys.6.1.005}{SciPost Phys. {\bf 6}, 005 (2019)}.


\bibitem{id-17}
E. Ilievski and J. De Nardis, {\it Ballistic transport in the one-dimensional Hubbard model: The hydrodynamic approach}, 
\href{	\doi10.1103/PhysRevB.96.081118}{Phys. Rev. B {\bf 96}, 081118(R) (2017)}.

\bibitem{mbpc-18}
M. Mestyan, B. Bertini, L. Piroli, and P. Calabrese, {\it Spin-charge separation effects in the low-temperature transport of 1D Fermi gases},
\href{	\doi10.1103/PhysRevB.99.014305}{Phys. Rev. B {\bf 99}, 014305 (2019)}.

\bibitem{nt-20}
Y. Nozawa and H. Tsunetsugu, {\it Generalized Hydrodynamic approach to charge and energy currents in the one-dimensional Hubbard model},
\href{https://arxiv.org/abs/1910.02427}{arXiv:1910.02427}.      

\bibitem{agarwal2018fast}
K. Agarwal, R. Bhatt, and S. Sondhi, {\it Fast preparation of critical ground states using superluminal fronts}, 
\href{\doi10.1103/PhysRevLett.120.210604}{Phys. Rev. Lett. {\bf 120}, 210604 (2018)}.

\bibitem{bernien2017probing}
 H. Bernien, S. Schwartz, A. Keesling, H. Levine, A. Omran, H. Pichler, S. Choi, A. Zibrov, M. Endres, M. Greiner, V. Vuletic, and M. Lukin, 
 {\it Probing many-body dynamics on a 51-atom quantum simulator}, 
\href{https://doi.org/10.1038/nature24622}{Nature {\bf 551}, 579 (2017)}.



\end{thebibliography}
\end{document}